\documentclass[twocolumn]{aastex631}

\newcommand{\eb}{\begin{equation}}
\newcommand{\ee}{\end{equation}}
\newcommand{\masyr}{mas yr$^{-1}$}
\newcommand{\gbp}{$G_{\rm BP}$}
\newcommand{\grp}{$G_{\rm RP}$}
\definecolor{rkka}{RGB}{219,66,32}

\shorttitle{Quasar proper motions and multiplicity}
\shortauthors{Makarov \& Secrest}

\begin{document}

\title{Quasars with Proper Motions and the Link to Double and Multiple AGNs}

\correspondingauthor{Valeri V. Makarov}
\email{valeri.makarov@gmail.com}

\author[0000-0003-2336-7887]{Valeri V. Makarov}
\affiliation{U.S. Naval Observatory, 3450 Massachusetts Ave NW, Washington, DC 20392-5420, USA}

\author[0000-0002-4902-8077]{Nathan J. Secrest}
\affiliation{U.S. Naval Observatory, 3450 Massachusetts Ave NW, Washington, DC 20392-5420, USA}

\begin{abstract}
Gaia used a large sample of photometrically selected active galactic nuclei (AGNs) and quasars to remove
the residual spin of
its global proper motion system in order to achieve a maximally inertial
reference frame. A small fraction of these reference objects have statistically significant astrometric proper motions in Gaia EDR3. We compile a source
sample of $105,593$ high-fidelity AGNs with accurate spectroscopically determined redshifts above 0.5 from
the SDSS and normalized proper motions below 4. The rate of genuinely perturbed proper motions is at least 0.17\%. A smaller high completeness sample of 152 quasars with excess proper motions at a confidence level of 0.9995 is examined in detail. Pan-STARRS images  and Gaia-resolved pairs reveal that
29\% of the sample are either double sources or gravitationally lensed quasars.
An Anderson--Darling test on parameters of a smaller high-reliability sample and their statistical controls reveals 17 significant factors that favor multiplicity and multi-source structure as the main cause of perturbed astrometry. Using a nearest neighbor distance statistical analysis and counts of close companions in Gaia on a much larger initial sample of AGNs, an excess of closely separated sources in Gaia is detected. At least 0.33\% of all optical quasars are genuinely double or multiply imaged. We provide
a list of 44 candidate double or multiple AGNs and four previously known gravitational lenses. Many proper motion quasars may be more closely separated, unresolved doubles exhibiting the
variability imposed motion (VIM) effect, and a smaller fraction may be chance alignments with foreground stars causing weak gravitational lensing.
\end{abstract}

\section{Introduction} \label{section: Introduction}
Quasars and point-like active galactic nucleus (AGN) sources have a special role in the construction of the fundamental optical reference frame. The
Gaia space astrometry mission \citep{2016A&A...595A...1G} provides, by far, the most accurate and densely populated celestial reference frame (CRF), called Gaia-CRF \citep{2021A&A...649A...1G}. However, the condition equations of space astrometry are intrinsically invariant with respect to six coordinate transformations, namely, a rigid rotation of the global coordinate system and a constant spin of the proper motion system. The former ambiguity can be removed in different ways because the orientation of the coordinate triad is a technical issue and a matter of convenience and convention. It was natural to align Gaia-CRF with the International Reference Frame (ICRF), which is based on VLBI phase-reference measurements of a few thousand radio-loud quasars \citep{2020yCat..36440159C} and realizes the International Celestial Reference System (ICRS). Determination and removal of the residual spin is, on the other hand, a more delicate subject that requires great care. This calculation uses a much greater number of mid-infrared (MIR)-identified quasars \citep{2015ApJS..221...12S} also observed by Gaia \citep{2018A&A...616A..14G} and is based on the assumption that quasars are, in general, in the ``rest frame'' of the large-scale universe and therefore define a non-rotating inertial frame. Owing to the remarkable progress both in the number of observed quasars and the measurement precision achieved in the latest Gaia data releases DR2 and EDR3 \citep{2018A&A...616A...1G,2021A&A...649A...1G}, we are entering a phase when this basic assumption can be realistically tested.

As quasars generally reside in galaxies, they have peculiar motions that can be several hundreds of kilometers per second relative to the Hubble flow \citep[e.g.,][]{2016AJ....152...50T}. The resulting astrometric proper motions should be negligible for most of them because of the great distances separating the sources from the observer. For a standard flat $\Lambda$CDM cosmology with $H_0=70$~km~s$^{-1}$~Mpc$^{-1}$ and $\Omega_\mathrm{m}=0.3$, one milliarcsecond subtends 8~pc at a typical quasar redshift of $z=1$, so a quasar with a peculiar velocity of 100~km~s$^{-1}$ will have an intrinsic proper motion of $\sim0.01$~$\mu$as~yr$^{-1}$, three orders of magnitude below even the most precise proper motions available from Gaia.

The situation may be different if we consider collective proper motion patterns, i.e., sky-correlated proper motion fields. These may emerge from purely astrometric errors of systematic or random origin \citep{2012AJ....144...22M, 2022ApJ...927L...4M}. At the 1 $\mu$as level of accuracy and beyond, a range of physical (relativistic) and cosmological phenomena become observable as discussed by \citet{2010IAUS..261..345M}. One of these phenomena, namely, the secular aberration drift caused by the Galactic acceleration of the solar barycenter ($\sim5$~$\mu$as~yr$^{-1}$), has been successfully estimated using Gaia EDR3 data \citep{2021A&A...649A...9G}. Other issues of great importance shall await the future Gaia data releases or future space missions \citep{2021FrASS...8....9K, 2022ApJ...927L...4M}. The main technical problem is how to separate instrumental sky-correlated errors from the genuine signals from the sky. Obviously, a genuine spin of the quasar ensemble, which may be caused by rotation of the universe, for example, is impossible to detect because it would have been removed in the alignment of the proper motion system. With the data in hand, we might expect that deviations of individual quasar proper motions from zero are of instrumental origin.

Quasars are not perfect sources of radiation for precision astrometry \citep{2012MmSAI..83..952M}. The closer AGNs at redshifts $z\lesssim 0.5$ often have resolved host galaxies associated with them, which are, in general, asymmetric to some degree. At higher redshift, double quasar systems are likely to exhibit astrometric variance, due either to the variability-induced motion (VIM) astrometric effect \citep[VIM;][]{2003A&A...399.1167P,2009aaat.book.....P,2016ApJS..224...19M} on the apparent photocenter of the combined system \citep[e.g.,][]{2020ApJ...888...73H}, or to source incompleteness in Gaia data for small separations \citep[$\lesssim 2\arcsec$;][]{2021A&A...649A...5F}. Indeed, unexplained astrometric variance has been successfully used to find double quasars \citep{2021NatAs...5..569S}, and cosmological simulations predict that the incidence of dual AGNs at moderate redshift is of order one to a few percent \citep[for a review, see][]{2019NewAR..8601525D}. At a much lower rate, quasars are gravitationally lensed by foreground galaxies with doubly or multiple-imaged configurations \citep{2019A&A...622A.165D}. The lensed images of currently known systems are mostly packed within a few arcseconds. These features can obviously produce position offsets in astrometry but not necessarily proper motion perturbations because they are long-term stable. Proper motion perturbations are more likely to arise due to differential variability between close sources, and may also arise due to the motion of a relativistic jet, as in the case of PKS~0119+11 \citep{2021AA...651A..64L}.

In this paper, we carefully estimate the rate of excess proper motions of moderate redshift quasars and explore the likely reasons for apparent proper motions in these objects. We use the mid-IR AGN (hereafter ``MIRAGN'') catalog of \citet{2015ApJS..221...12S}, cross matched to Gaia EDR3. Smaller test samples of quasars with spectroscopically determined redshifts from the Sloan Digital Sky Survey \citep[SDSS][]{2020ApJS..250....8L} and statistically significant proper motions are generated and the available Pan-STARRS images \citep{2010SPIE.7733E..0EK} are inspected for source structures and double sources. We also analyze the near-neighbor distance statistics of resolved companions to a larger sample of 0.632 million quasars present in Gaia EDR3 within $11\arcsec$ to confirm the presence of real binary AGNs and estimate their overall rate. Weak gravitational lensing by foreground stars as an alternative cause of perturbed proper motions is briefly discussed as well.

\section{Methodology} \label{section: Methodology}
\subsection{Astrometric quasar sample}
We cross match the catalog of 1.4~million MIRAGNs from \citet{2015ApJS..221...12S} to the Gaia EDR3 catalog, using a match tolerance of $0\farcs5$ for reliability, as later selection on proper motion preferentially selects on the very small fraction of contaminant stars in the MIRAGN catalog. This produced 621,946 matches, 551,482 of which have proper motion measurements. As Gaia's astrometric processing is designed for compact and unresolved objects, AGNs hosted in extended galaxies at low redshift may have spurious astrometry. To avoid this, we remove objects at low redshifts (Section~\ref{subsec: QC}) by matching the Gaia counterpart coordinates to to the SDSS specObj-dr16.fits table,\footnote{\url{https://www.sdss.org/dr16/spectro/spectro_access/}} allowing only spectra with \texttt{ZWARNING==0} or \texttt{ZWARNING==4}, the latter of which can happen for spectra with broad lines.\footnote{\url{https://www.sdss.org/dr16/spectro/caveats/\#zstatus}} To ensure spectroscopic fiber coverage of the Gaia counterpart, we allow BOSS spectra within $1\arcsec$ and SDSS spectra within $1\farcs5$. This produced matches for 126,343 objects out of the full sample of 621,946 MIRAGN-Gaia matches.

\subsection{Astrometric corrections and quality control} \label{subsec: QC}
\begin{figure}
    \includegraphics[width=\columnwidth]{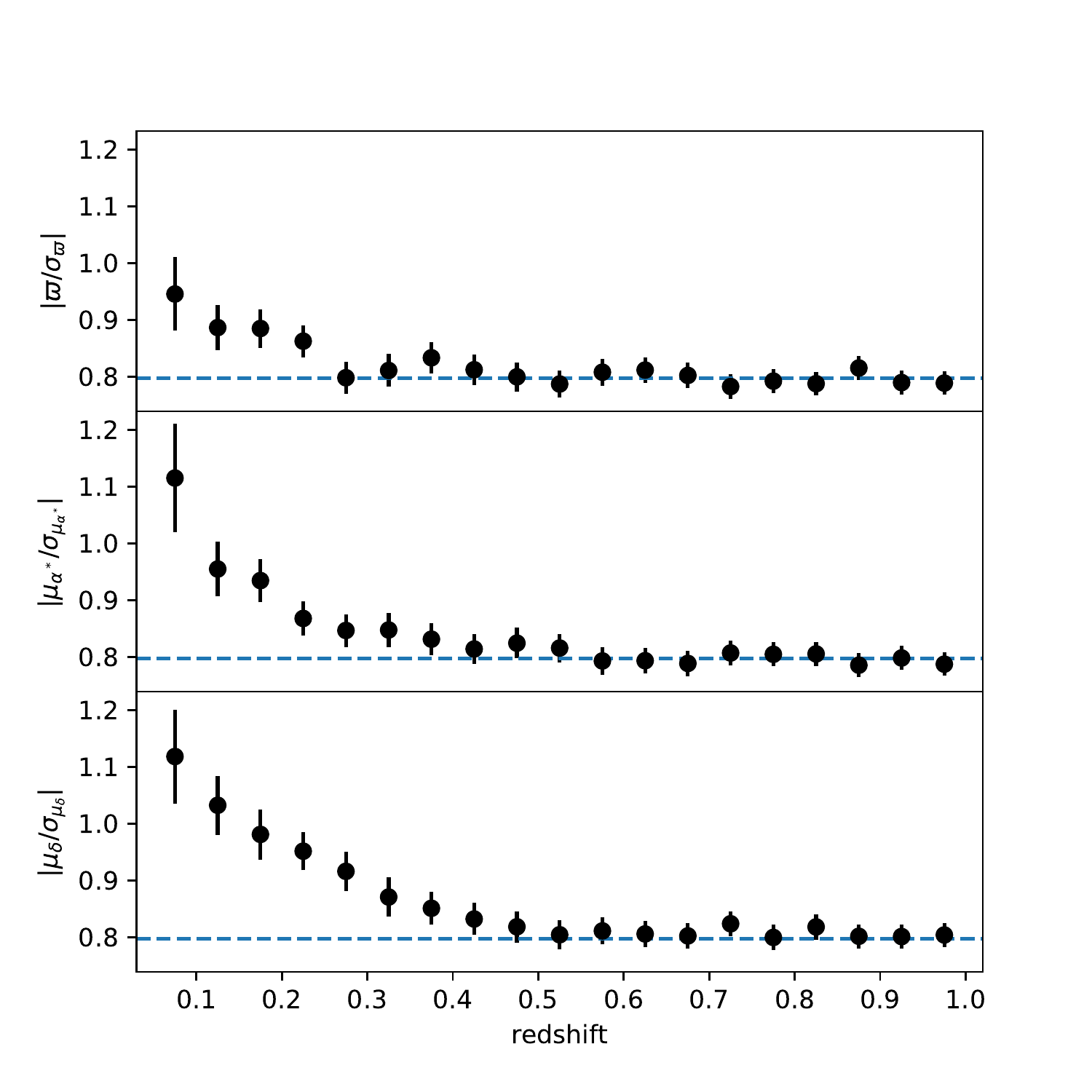}
    \caption{Distributions of the absolute, corrected, error-normalized parallaxes (top panel) and proper motions (middle panel for R.A.; bottom panel for decl.), binned by redshift, demonstrating the need for the redshift cut $z>0.5$. The dashed lines show the expectation for normally-distributed values with a mean of zero and sigma of unity ($\sqrt{2/\pi}$). The error bars are 2 times the standard error of the mean within the bin.}
    \label{fig:npara_npmra_npmdec_vs_z}
\end{figure}

We find that, for redshifts greater than $z>0.5$, the distributions of error-normalized parallax and proper motions are Gaussian, but require minor zero-point and error corrections. We determined these correction by selecting objects with error-normalized values less than 4 and iteratively adding an overall zeropoint offset along with a multiplicative error correction factor until the distributions reach a mean of zero and standard deviation of unity. For parallax, this zero-point correction is $+19.7$~$\mu$as, with errors multiplied by 1.053. For proper motion in R.A.\ ($\alpha$), these values are $-3$~$\mu$as yr$^{-1}$ and 1.056. For proper motion in decl.\ ($\delta$), these values are $+2$~$\mu$as yr$^{-1}$ and 1.064. The significances of these zero-point offsets are $25\sigma$, $4.5\sigma$, and $2.6\sigma$, respectively. Our empirically determined error correction factors are consistent with the standard deviation of normalized proper motions of 1.063 determined by \citet{2021A&A...649A...9G}, as well as with the offsets determined in \citet{2022A&A...660A..16S}. We show the absolute values of these corrected, error-normalized quantities as a function of redshift in Figure~\ref{fig:npara_npmra_npmdec_vs_z}, demonstrating that AGNs with redshifts below $0.5$ have systematically elevated significances. 

\begin{figure}[htbp]
  \centering
  \plotone{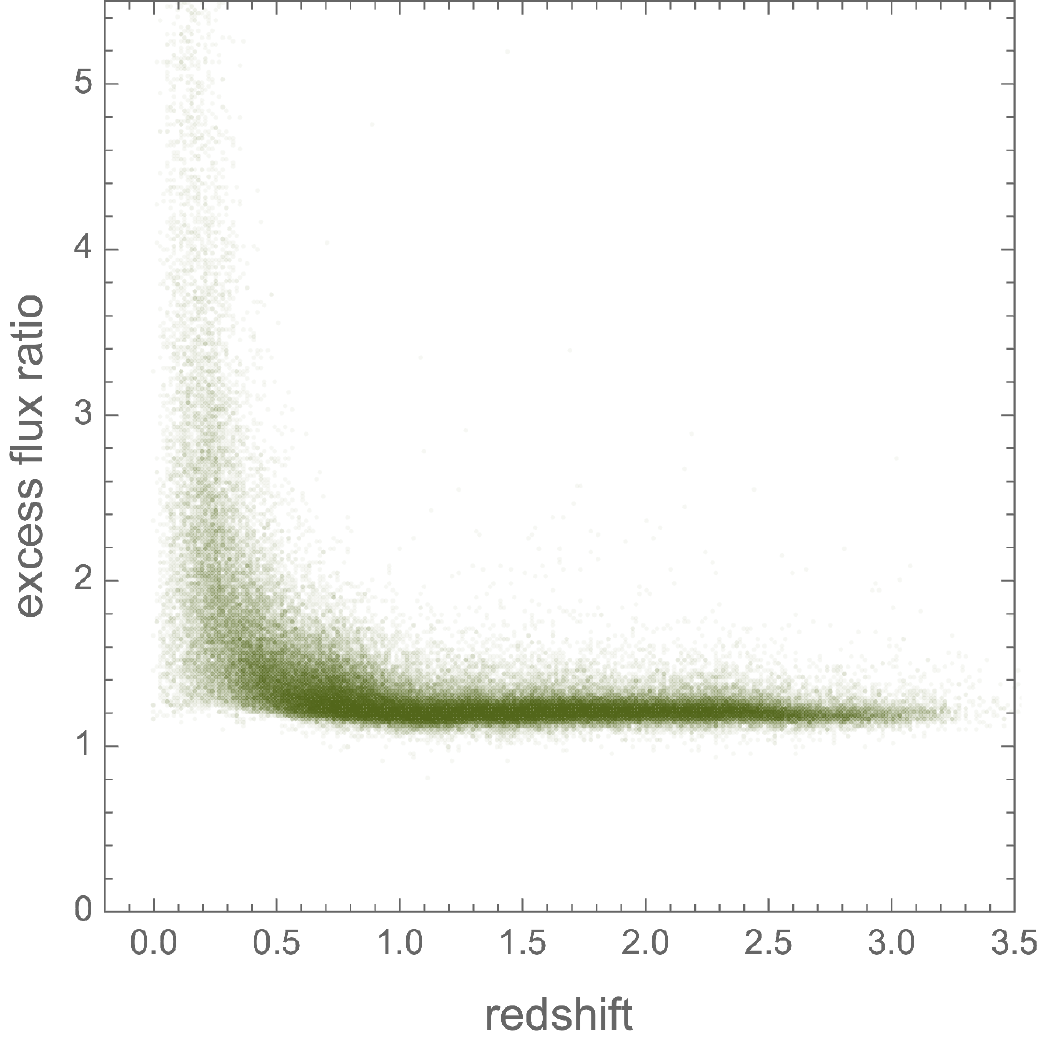}
\caption{Photometric excess ratio (\texttt{phot\_bp\_rp\_excess\_factor}) values from Gaia EDR3 versus spectroscopic redshifts from SDSS for cross-matched MIRAGNs. \label{evsz.fig}}
\end{figure}

The origin of astrometric deterioration at $z<0.5$ is illustrated by analysis of the photometric excess parameter (\texttt{phot\_bp\_rp\_excess\_factor}) provided in the Gaia EDR3 catalog as a quality check. This quantity is computed as the flux ratio (Flux\_BP+Flux\_RP)/Flux\_G \citep[][see also \url{https://gea.esac.esa.int/archive/documentation/GEDR3/Catalogue_consolidation/chap_cu9val/sec_cu9val_942/ssec_cu9val_942_photometry.html}]{2021A&A...649A...3R}.  Because of the different methods used in EDR3 to estimate the broad-band magnitudes $G$ and the narrower \gbp\ and \grp\ magnitudes, the ratio of the corresponding fluxes may deviate upward significantly from the most common value slightly above 1. The Gaia EDR3 data reduction pipeline was not tailored for extended objects and any bright optical structure makes the astrometric results inaccurate, including proper motions. This is seen in Figure \ref{evsz.fig} where the excess flux ratio (called $\cal F$ in the following for brevity) is plotted against spectroscopic redshifts. While the majority of more distant and luminous quasars tightly group around a well-defined lower envelope at ${\cal F}\simeq 1.1$, the closer AGNs show a dramatic increase both in the median and dispersion values. This is definitely caused by the more prominent presence of host galaxies in the images of MIRAGNs at $z<0.5$. Intrinsically more luminous AGNs have a higher ratio of core$/$galaxy flux, so that the contribution of underlying host structures tapers off with $z$. Thus, our initial selection includes $105,596$ MIRAGNs with $z>0.5$.

\begin{figure*}
    \includegraphics[width=\textwidth]{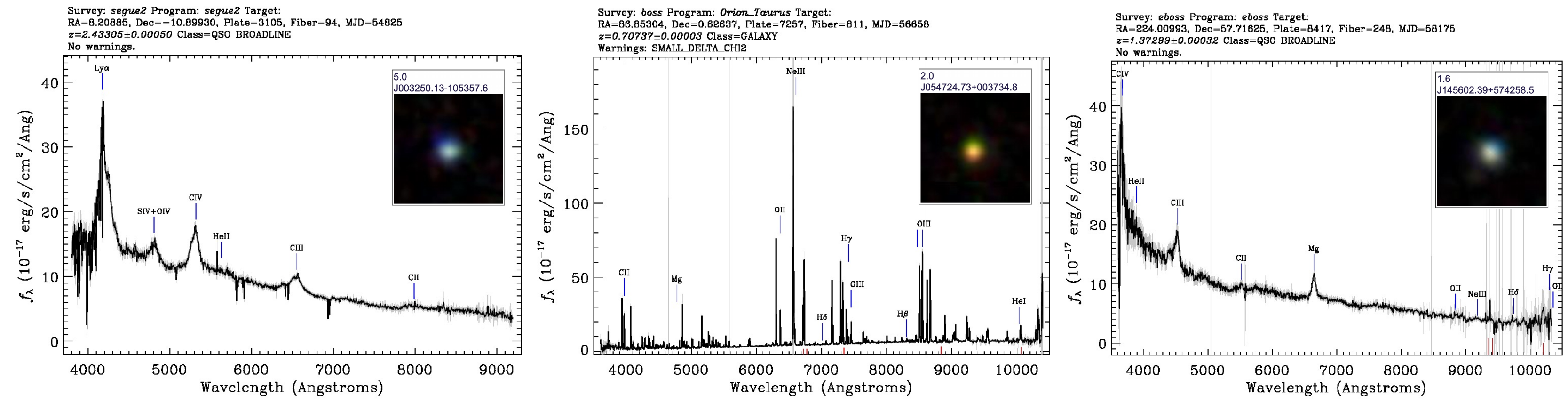}
    \caption{SDSS/BOSS spectra of three MIRAGNs with significant ($>5\sigma$) parallaxes identified in this work, with their DR16 $gri$ $12\arcsec\times12\arcsec$ thumbnails inset. The value of the parallax, in mas, is given in the header of each thumbnail above the SDSS identifier. The middle object, WISEAJ054724.73+003734.8, is identified in SIMBAD as a young stellar object, explaining its parallax, and its spectrum is not that of a quasar. The other two objects, however, have quasar spectra and are not cross-identified with any stellar type.}
    \label{fig:parallax_objects}
\end{figure*}

After zero-point and error correction, we find 7 objects with normalized proper motions greater than 4, out of 105,596 AGNs (0.007\%). Four of these have normalized parallaxes less than 4.7, which is generally consistent with expectations for normally distributed random data, given the sample size. However, three objects have normalized parallaxes greater than 5, which is significant. One of these three objects, WISEAJ054724.73+003734.8, is identified in SIMBAD as a young stellar object, but the other two objects are quasars with unexplained parallax. We show these objects and their SDSS/BOSS spectra in Figure~\ref{fig:parallax_objects}. We trim these three objects from our sample, leaving 105,593 objects.

\subsection{Statistical significance of proper motions} \label{rate.sec}
\begin{figure}
    \includegraphics[width=\columnwidth]{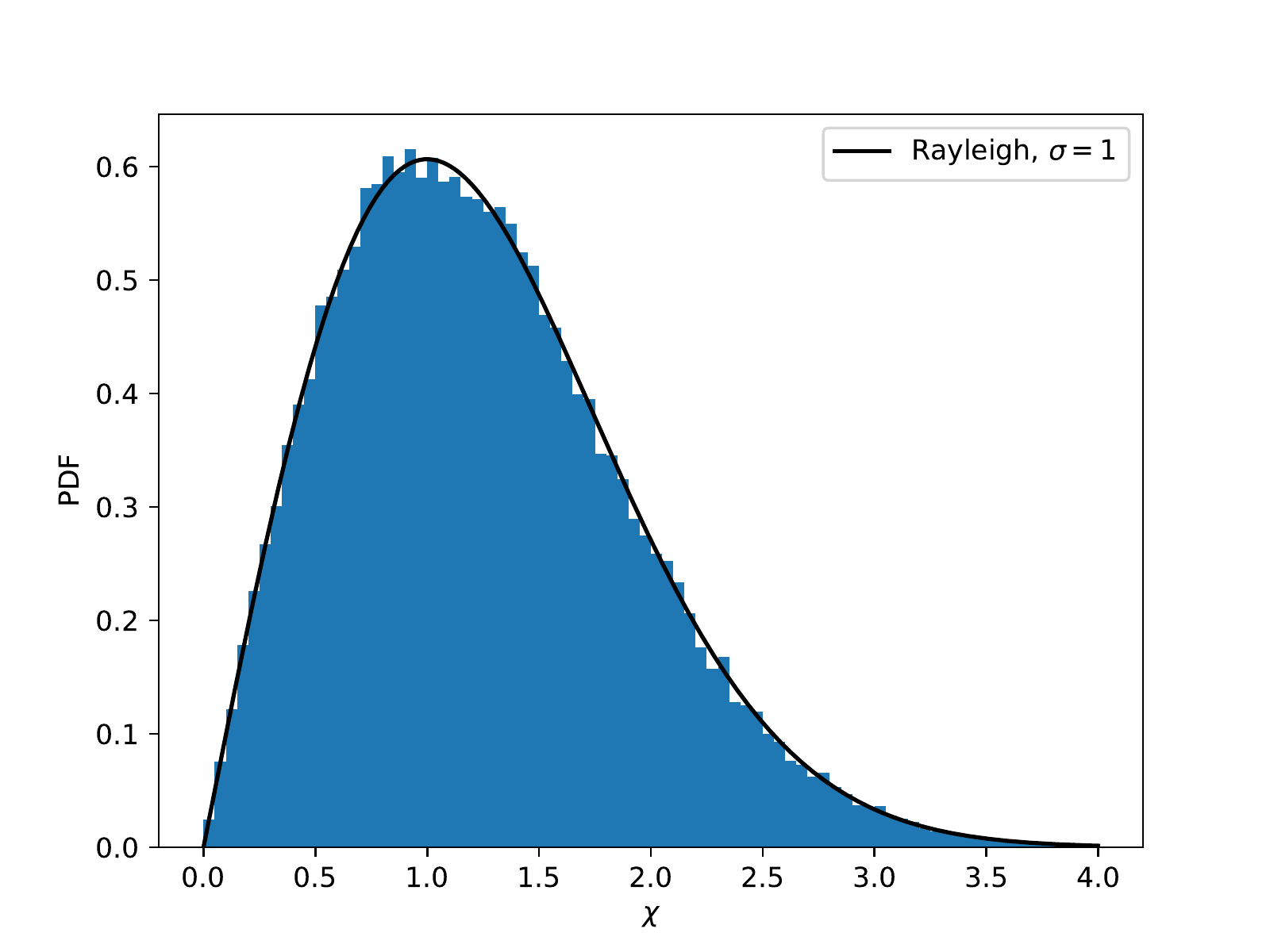}f
    \caption{Distribution of proper motion $\chi$ values (Equation~\ref{eq: chi2}) after correcting the uncertainties and zero-point offset.}
    \label{fig:pmchi}
\end{figure}

To find objects with significant proper motions, we calculate $\chi^2$ for the error-normalized proper motions, including the correlation term, as:

\begin{equation} \label{eq: chi2}
\chi^2 = \frac{1}{1 - \rho^2} \left[ \left( \frac{\mu_{\alpha^*}}{\sigma_{\mu_{\alpha^*}}} \right)^2 + \left( \frac{\mu_{\delta}}{\sigma_{\mu_{\delta}}} \right)^2 - 2 \rho \frac{\mu_{\alpha^*} \mu_{\delta}}{\sigma_{\mu_{\alpha^*}} \sigma_{\mu_{\delta}}} \right],
\end{equation}

\noindent where $\mu_{\alpha^*} = $~\texttt{pmra}, $\mu_{\delta} = $~\texttt{pmdec} are the proper motions in R.A.\ and decl.\ provided in the Gaia catalog, and $\rho=$~\texttt{pmra\_pmdec\_corr} is the correlation between them. For objects with accurate uncertainties and no intrinsic proper motions, the square-root of $\chi^2$ follows a Rayleigh distribution with $\sigma=1$. We find that, for values of $\chi$ below 4, the distribution of $\chi$ is well-fit with this Rayleigh distribution (Figure~\ref{fig:pmchi}), with a reduced chi-squared of 1.4 (dof~$=79$).

With the error-normalized proper motions $\chi$ following the expected Rayleigh distribution, determining the cut in $\chi$ for a reliable sample of AGNs with apparent proper motions is straightforward. The presence of statistically significant quasar proper motions is best seen as the excess of objects with large $\chi^2$ values compared to the expected rates. The cumulative distribution function (CDF) of the $\chi^2(2)$ distribution with two degrees of freedom for this bivariate statistic) can be used to calculate the probability of this value being greater than a certain limit. For example, CDF$_{\chi^2(2)}(9)=0.988891$, so only 1.111\% of the source sample is expected to have $\mu/\sigma_\mu>3$. For a given confidence level $\Phi$ sufficiently close to 1, the expected rate of outliers is $1-\Phi$, which can be compared with the actual rate of objects whose $\chi$ is greater than $\sqrt{{\rm CDF}^{-1}_{\chi^2(2)}(\Phi)}$. At $\Phi=0.998$, we expect to find 211 quasars out of $105,593$ with $\mu/\sigma_\mu>3.52551$ while, in reality, 326 such objects are present. Therefore, 115 objects, or 0.11\% of the source sample, have elevated proper motions not accounted for by the assumed statistical distribution. 

\begin{figure}[htbp]
  \centering
  \plotone{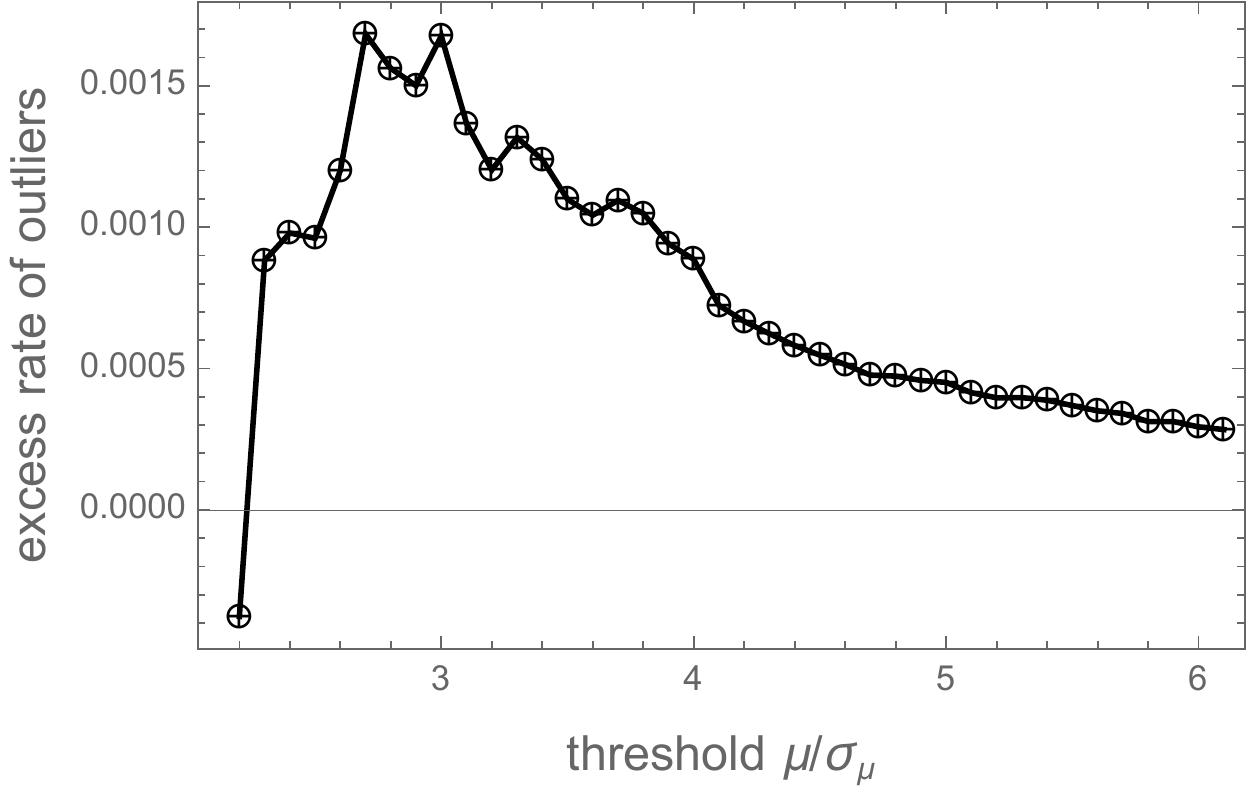}
\caption{The excess rate of outliers with normalized proper motions $\chi$ above specific threshold values for the source sample of $105,593$ high-fidelity quasars. \label{rate.fig}}
\end{figure}

Following these lines, we computed the expected and the actual numbers of outliers for a grid of threshold $\chi$ values. The result is represented in Figure \ref{rate.fig} as a sample rate of the excess, i.e., the difference between the latter and the former divided by the total number of objects in the source sample. The peak of this function indicates that up to about 0.0017 of the source sample have excessive proper motions as measured by Gaia EDR3. Most of them, however, have rather weakly measured proper motions, with $\chi$ between 2.8 and 3.0. 

\section{Quasars with Proper Motions}
Given the low rate of objects with statistically significant proper motions in the initial source sample, a blind search of physically perturbed objects would be extremely inefficient. Our aim is now to generate much smaller test samples of objects maximizing the rate of genuinely perturbed motions. These are strongly diluted with regular statistical outliers (random flukes) at moderate $\chi=\mu/\sigma_\mu$ levels. Therefore, taking a higher threshold $\chi$ value dramatically increases the relative rate of statistical outliers. 

\begin{deluxetable*}{lclrrr} \label{tab:reliable_ad}
\tablecaption{
Parameters with significant ($p<0.05$) differences between the 47 objects from the reliable sample and their matched controls}
\tablehead{\colhead{parameter} & \colhead{unit} & \colhead{meaning} & \colhead{sample} & \colhead{control} & \colhead{$p$-value} }
\startdata
RUWE &   & single point source fit quality & 1.7(0.2) & 1.016(0.002) & $<0.001$ \\
nearest neighbor & arcsec & offset from nearest EDR3 neighbor & 12(2) & 18.1(0.4) & $<0.001$ \\
EDR3/specObj offset & arcsec & EDR3 offset from spectroscopic fiber & 0.10(0.01) & 0.064(0.003) & $<0.001$ \\
$G_\mathrm{BP}-G_\mathrm{RP}$ & mag & optical color & 0.86(0.05) & 0.593(0.009) & $<0.001$ \\
BP/RP excess factor & mag & source extent, binarity, crowding & 1.51(0.04) & 1.251(0.005) & $<0.001$ \\
\texttt{ipd\_frac\_odd\_win} & & nearby source contamination & 0.7(0.5) & 0.07(0.03) & $<0.001$ \\
$\varpi/\sigma_\varpi$ & & parallax significance & -0.8(0.4) & -0.01(0.04) & $<0.001$ \\
\texttt{astrometric\_primary\_flag} & boolean & astrometrically well-behaved source & 0.02(0.02) & 0.91(0.01) & $<0.001$ \\
\texttt{astrometric\_params\_solved} & & \texttt{pseudocolour} estimated$^\dagger$ & 54(5) & 36.4(0.8) & $<0.001$ \\
\texttt{astrometric\_excess\_noise} & mas & unexplained astrometric variance  & 2.0(0.3) & 0.25(0.02) & $<0.001$ \\
\texttt{astrometric\_excess\_noise\_sig} &  & significance of \texttt{astrometric\_excess\_noise} & 40(20) & 0.40(0.03) & $<0.001$ \\
\texttt{ipd\_frac\_multi\_peak} & & percent of windows with multiple sources  & 7(2) & 0.10(0.02) & $<0.001$ \\
\texttt{astrometric\_n\_bad\_obs\_al} & & number of downweighted observations & 4.0(1.0) & 1.85(0.08) & 0.007 \\
$G-W2$ & mag & optical to mid-IR color & 5.3(0.1) & 4.98(0.03) & 0.010 \\
\texttt{astrometric\_sigma5d\_max} & mas & max error of 5 astrometric parameters & 0.64(0.09) & 0.44(0.02) & 0.023 \\
neighbors ($<30\arcsec$) & & number of EDR3 neighbors within $30\arcsec$ & 1.4(0.2) & 1.04(0.06) & 0.028 \\
redshift & & spectroscopic redshift & 1.46(0.08) & 1.29(0.03) & 0.035
\enddata
\tablecomments{The means and standard errors of the means are given, along with the $p$-value calculated using the Anderson-Darling test. References for the meaning of these parameters are the Gaia online documentation and \citet{2012AA...538A..78L, Lindegren21,Riello21}.\\
$^\dagger$ The sample and controls have either \texttt{astrometric\_params\_solved}~$=31$ or \texttt{astrometric\_params\_solved}~$=95$, respectively, indicating either a five-parameter or six-parameter astrometric solution. The sixth parameter is \texttt{pseudocolour}, which is solved when the source has an issue such as being in a crowded field \citep{Lindegren21}.} 
\end{deluxetable*}

We produced two samples for this study, using two threshold values of confidence: a ``high reliability'' sample, which is expected to have zero objects with apparent proper motions due to random chance, and a ``high completeness'' sample, which has a much larger number of objects with real apparent proper motions, at the cost of reliability. We compare the high reliability sample with a control sample of objects without apparent proper motions, in order to determine parameter differences that may reveal the cause of apparent proper motion in quasars. The high completeness sample serves as an important sample for follow-up studies, but we manually inspect this sample in order to remove as many false positives as possible, and discuss their overall properties. 

\subsection{High reliability sample} \label{subsec: high reliability sample}
The number of false positives as a function of $\chi$ cut is given by multiplying the sample size $n$ by the survival function. For 105,593 objects, a cut of $\chi > 5$ gives an expected number of false positives of $\sim0$. We find 47 objects in excess of this criterion (Table 3).

To understand the reasons for apparent proper motion in quasars, we produced a control sample of objects from the same parent sample but with $\chi < 4$. We matched on Gaia~$G$ magnitude and S/N, which is good proxy for sensitivity to astrometric motion, as well as choosing the closest controls on the sky meeting these match tolerances, in order to negate position-dependent motion sensitivity differences. We found that, by matching to within $\pm0.5$~mag and 0.2~dex in S/N, we were able to produce 10 unique controls for each object in the proper motion sample (470 total controls). An Anderson-Darling test gives $p>0.25$ for Gaia~$G$ magnitude and S/N. 

We then compared the sample and controls by performing an Anderson--Darling test on parameters of potential astrophysical or systematic relevance, such as photometric color, astrometric fit quality, or number of nearby neighbors. We find 17 parameters with $p < 0.05$, listed in Table~\ref{tab:reliable_ad}. There are a number of causes for apparent proper motion suggested by these differences:
\begin{enumerate}
\item Multi-sources: the presence of secondary sources is indicated by the very high significance ($p<0.001$) of differences in the Renormalised Unit Weight Error (RUWE), Gaia EDR3/specObj offset (indicating a potential shared centroid for multiple sources in lower resolution SDSS imaging), BP/RP excess factor (an indicator of source extent and unresolved multiplicity), \texttt{ipd\_frac\_odd\_win} (an indicator of contamination by a nearby source), \texttt{astrometric\_excess\_noise} \citep[a measure of unexplained astrometric variance in the source, often attributed to dual AGNs; e.g.][]{2019ApJ...885L...4S,2020ApJ...888...73H} and its significance, and possibly \texttt{astrometric\_n\_bad\_obs\_al}, the number of CCD transits strongly down-weighted in the astrometric solution. 

\item Environment: there is also strong evidence that the nearest Gaia EDR3 neighbors for the sample are considerably closer than for the controls, potentially indicating spurious contamination by foreground stars or nearby sources resolved by Gaia (see Sec. \ref{nn.sec}). However, the mean number of neighbors within $30\arcsec$ is only slightly higher for the sample than the controls, indicating either that the sample objects are in slightly denser environments than for the controls, or that the fraction of sample objects with close contaminants is small.  

\item Source properties: sample objects have $\sim0.3$~mag redder optical color ($G_\mathrm{BP}-G_\mathrm{RP}$) than the controls, on average, and their optical-to-mid-IR color ($G-W2$) is 0.3~mag redder on average, although this latter difference is less significant ($p=0.010$). These differences are not attributable to the somewhat higher mean redshift of the sample compared with its controls (1.46 versus 1.29, although this difference is only marginally significant at $p=0.035$), which induces only a $+0.05$~mag difference in $G_\mathrm{BP}-G_\mathrm{RP}$ and a $-0.15$~mag difference in $G-W2$, estimated from the full sample of MIRAGN-EDR3 matches. This suggests that quasars with apparent proper motions are often redder than those without, indicating a potential role of AGN obscuration.

\item Unknown: the sample differs from its controls in \texttt{astrometric\_primary\_flag}, \texttt{astrometric\_params\_solved},\\ and \texttt{astrometric\_sigma5d\_max}, which do not have clear explanations, but may be due to requiring more complex astrometric solutions for multi-sources and difficulties with obtaining spectrophotometric measurements. The sample also has systematically negative normalized parallaxes, which are nonphysical but may again be due to complex astrometry.
\end{enumerate}

With these considerations, source multiplicity is the clearly favored explanation for apparent proper motion in quasars. We explore this in the next sections, but for the reader interested in conducting a similar control analysis, we provide the high reliability sample as a table in the Appendix.

\subsection{High completeness sample}
To produce a test sample small enough to be studied on an object-by-object basis, we choose a confidence level of $\Phi=0.9995$ and select objects above the corresponding threshold $\mu/\sigma_\mu>3.89895$. This cut leaves only 152 objects. The expected number of regular statistical outliers is 53. Therefore, 99 quasars, or 65\% of the test sample, are objects with genuinely perturbed proper motions. 

In order to find out the origin of the excessive proper motions in the test sample of 152 quasars with redshifts greater than 0.5, we manually reviewed all the images available through the online Pan-STARRS cutout service.\footnote{\url https://ps1images.stsci.edu/cgi-bin/ps1cutouts} The footprint of Pan-STARRS $3\pi$ survey (declination $>-30\degr$) includes all 152 quasars. Approximately two-thirds of the images do not show any signs of peculiarities, exhibiting unresolved morphologies in line with QSOs. Although one-half of this proportion is consistent with the expected number of statistical outliers, the results from the high-reliability sample (Section~\ref{subsec: high reliability sample}) suggests that many of these objects have sub-arcsecond companions, resolved by Gaia (up to $0\farcs4$ angular resolution) but not by Pan-STARRS ($1\farcs1$ median seeing in the $i$ band). Approximately one-third of the sample, however, show unusual features, mostly closely separated companions of fainter magnitudes within $2\farcs5$ of the target AGNs. All these objects are listed in Table \ref{cand.tab} with notes describing the features.  For each resolved or suggested companion, a pair of numbers in the notes specifies the angular separation  and position angle (north through east). These values are computed from the astrometric data in the Gaia EDR3 catalog for 21 resolved sources, or roughly estimated by eye from Pan-STARRS images.

\startlongtable
\begin{deluxetable*}{C CC CrC cCC l}
\tabletypesize{\footnotesize}
\tablecaption{Candidate double or lensed AGNs.\label{cand.tab}}\tablehead{
\colhead{\rm MIRAGN~ObjID}  & \colhead{\rm RA}   & \colhead{\rm decl.} & \colhead{$z$} & \colhead{$\chi$} & \colhead{$G$} & \colhead{\rm Res.} & \colhead{$\rho$} & \colhead{\rm PA} & \colhead{\rm notes} \\ [-0.2cm]
  	& \colhead{(\arcdeg)}	& \colhead{(\arcdeg)}	&   &   & \colhead{(mag)} & &  \colhead{(\arcsec)} & \colhead{(\arcdeg)} & } 
\startdata
\text{J015705.90+111253.6} & 029.27458 & +11.21491 & 1.083 & 3.97 & 20.52 & 2 & 2.1 & 100 & \\
\text{J024634.09-082535.9} & 041.64212 & $-$08.42672 & 1.686 & 4.52 & 16.92 & 1 & 1.096 & 304.6 & 1 \\
\text{J033847.66+010057.9} & 054.69866 & +01.01602 & 0.953 & 4.88 & 20.13 & 2 & 1.0 & 305 & \\
\text{J075824.27+145752.5} & 119.60110 & +14.96457 & 2.568 & 10.14 & 19.26 &  &.    &.    & 2 \\
\text{J080559.23+490742.0} & 121.49682 & +49.12836 & 1.008 & 13.49 & 18.53 & 1 & 0.514 & 225.9 & 3 \\
\text{J081331.28+254503.1} & 123.38030 & +25.75086 & 1.510 & 8.37 & 16.34 & 1 & 0.817 & 76.3 & 4 \\
\text{J083531.06+252808.7} & 128.87949 & +25.46910 & 1.024 & 5.49 & 19.69 & 2 & 0.9 & 110 & \\
\text{J085448.87+200630.6} & 133.70365 & +20.10851 & 0.778 &  4.00 & 14.54 &  &.   &.    & 5 \\
\text{J091831.59+110653.1} & 139.63161 & +11.11474 & 1.646 &  5.14 & 19.04 & 2 & 0.9 & 60 & \\
\text{J095122.59+263513.7} & 147.84404 & +26.58723.& 1.248 & 4.02 & 17.49 & 1 & 1.099 & 125.4 & 1 \\
\text{J095738.17+552257.7} & 149.40910 & +55.38271 & 0.901 & 10.61 & 17.24  & &   &.  &  6 \\
\text{J101051.15+570531.0} & 152.71311 & +57.09190 & 1.961 & 4.06 & 17.35 & 1 & 0.827 & 9.9 & \\
\text{J112948.61+133719.5} & 172.45259 & +13.62210 & 0.777 & 4.057 & 18.90 & 2 & 1.1 & 310 & \\
\text{J122016.86+112628.2} & 185.07031 & +11.44114 & 1.871 & 5.71 & 18.16 &  &  &  &  7 \\
\text{J122321.24+310313.8} & 185.83854 & +31.05381 & 0.885 & 5.06 & 19.15 & 2 & 1.5 & 135 & \\
\text{J123143.56+284749.7} & 187.93155 & +28.79716 & 0.859 & 5.78 & 16.73 & 2 & 0.7 & 140 & 8 \\
\text{J125617.97+584550.0} & 194.07498 & +58.76390 & 1.204 & 5.43 & 18.19 & 1 & 1.314 & 263.0 & \\
\text{J125631.35+330253.1} & 194.13063 & +33.04802 & 2.559 & 5.09 & 20.04 & 1 & 0.505 & 358.3 & \\
\text{J132559.51+144110.4} & 201.49802 & +14.68623 & 1.104 & 28.01 & 19.23 & 1 & 0.648 & 278.5 & \\
\text{J133127.36+322824.6} & 202.86401 & +32.47350 & 1.779 & 6.91 & 18.60 & 2 & 2.0 & 350 & \\
\text{J133512.10+052732.4} & 203.80054 & +05.45902 & 1.954 & 4.96 & 19.53 & 2 & 1.2 & 265 & \\
\text{J135346.06+183753.2} & 208.44195 & +18.63144 & 0.917 & 4.14 & 18.36 & 2 & 0.7 & 315 & 3 \\
\text{J135907.86+464419.0} & 209.78277 & +46.73863 & 0.595 & 4.04 & 19.08 & 2 & 1.2 & 315 & \\
\text{J140901.89+294633.5} & 212.25787 & +29.77597 & 1.585 & 9.83 & 18.85 & 1 & 0.907 & 307.2 & \\
\text{J143100.00-013141.7} & 217.75000 & -01.52827 & 0.829 & 9.41 & 19.05 &  &  &  &  2 \\
\text{J144034.78+441520.5} & 220.14495 & +44.25570 & 0.805 & 8.40 & 18.17 & 1 & 0.866 & 52.9 & \\
\text{J144444.64+223902.5} & 221.18606 & +22.65075 & 1.730 & 5.32 & 19.35 & 2 & 0.9 & 290 & \\
\text{J150251.19+144349.2} & 225.71330 & +14.73033 & 0.558 & 4.02 & 18.59 &  &  &  &  2 \\
\text{J150414.15+581611.8} & 226.05897 & +58.27008 & 1.036 & 4.03 & 19.31 & 2 & 0.8 &  & 9 \\
\text{J152024.51+211155.4} & 230.10204 & +21.19872 & 1.504 & 8.83 & 18.26 & 1 & 1.191 & 52.5 & \\
\text{J152902.83+384103.1} & 232.26182 & +38.68421 & 2.012 & 5.60 & 18.81 & 1 & 0.888 & 265.1 & \\
\text{J153038.05+545631.7} & 232.65863 & +54.94208 & 1.564 & 5.05 & 18.54 & 1 & 0.938 & 19.4 & \\
\text{J153509.63+082347.1} & 233.79019 & +08.39638 & 1.955 & 4.39 & 19.18 & 2 & 1.3 & 255 & 10\\
\text{J153959.23+320510.6} & 234.99682 & +32.08629 & 1.505 & 4.52 & 19.28 &  &  & & 2 \\
\text{J160614.71+230518.0} & 241.56123 & +23.08836 & 1.204 & 6.06 & 18.92 & 1 & 1.294 & 88.9 & \\
\text{J161827.73+505817.6} & 244.61552 & +50.97156 & 1.808 & 17.58 & 19.76 & 2 & 0.7 & 225 & \\
\text{J162207.39+545213.0} & 245.53078 & +54.87025 & 1.155 & 4.43 & 18.73 & 1 & 0.688 & 0.6 & \\
\text{J164258.81+394836.9} & 250.74504 & +39.81028 & 0.595 & 4.40 & 18.17 &  &  &  & 11 \\
\text{J172224.15+352019.2} & 260.60063 & +35.33867 & 1.126 & 5.69 & 19.84 & 1 & 1.181 & 169.4 & 1 \\
\text{J165043.44+425148.9} & 252.68102 & +42.86372 & 1.540 & 5.92 & 17.67 & 1 & 0.602 & 357.9 & \\
\text{J165713.02+303822.2} & 254.30423 & +30.63947 & 1.398 & 4.06 & 18.95 & 1 & 0.634 & 123.7 & \\
\text{J172758.48+564419.4} & 261.99371 & +56.73868 & 1.773 & 8.72 & 19.85 & 2 & 0.8 & 295 & \\
\text{J213932.19-011405.4} & 324.88405 & $-$01.23495 & 1.230 & 9.02 & 19.78 & 2 & 1.1 & 45 & \\
\text{J225738.51+204223.7} & 344.41044 & +20.70649 & 1.674 & 4.80 & 19.10 & 2 & 0.6 & 315 & \\
\text{J233522.50+320109.2} & 353.84383 & +32.01919 & 0.904 & 4.13 & 19.46 & 1 & 0.614 & 3.3 & \\
\text{J234330.59+043557.9} & 355.87760 & +04.59939 & 1.607 & 7.15 & 18.64 & 1 & 1.231 & 290.4 & \\
\text{J235422.48+195141.3} & 358.59369 & +19.86149 & 0.755 & 7.52 & 18.90 & 2 & 1.1 & 375 & \\
\enddata
 \tablecomments{Columns description: 1) MIRAGN object name from \citet{2015ApJS..221...12S}; 2) RA in degrees;
 3) Dec in degrees; 4) spectroscopic redshift from SDSS; 5) significance of Gaia EDR3 proper motion; 6) Gaia EDR3 mean $G$ magnitude; 7)
resolved in 1: Gaia EDR3, 2: Pan-STARRS images; 8) separation in arcseconds; 9) position angle in degrees 10) notes and references.}
\tablecomments{\\
1) known gravitational lens\\
2) unresolved companion or lensed image\\
3) possible gravitational lens\\
4) known gravitational lens HS 0810+2554\\
5) extended image, well studied candidate binary black hole OJ 287\\
6) 4C 55.17, radio-loud source, in ICRF3 (1), blazar\\
7) possible unresolved companions or lensed image\\
8) known BLL and blazar\\
9) two companions or lensed image\\
10) more distant companion resolved in Gaia EDR3 at $2.558\arcsec$, $358.1\degr$\\
11) 3C 345, radio-loud source, in ICRF3 (1), radio-optical offset (3), blazar\\
}
\tablerefs{(1) \citet{2020yCat..36440159C}; 
(2) \citet{2021AA...651A..64L}; (3) \citet{2017ApJ...835L..30M}; (4) \citet{2011yCat..35290091T};
(5) \citet{2006AJ....131.1934I}
}
\end{deluxetable*}
 
\begin{figure*}
\centering
\includegraphics[width=0.975\textwidth]{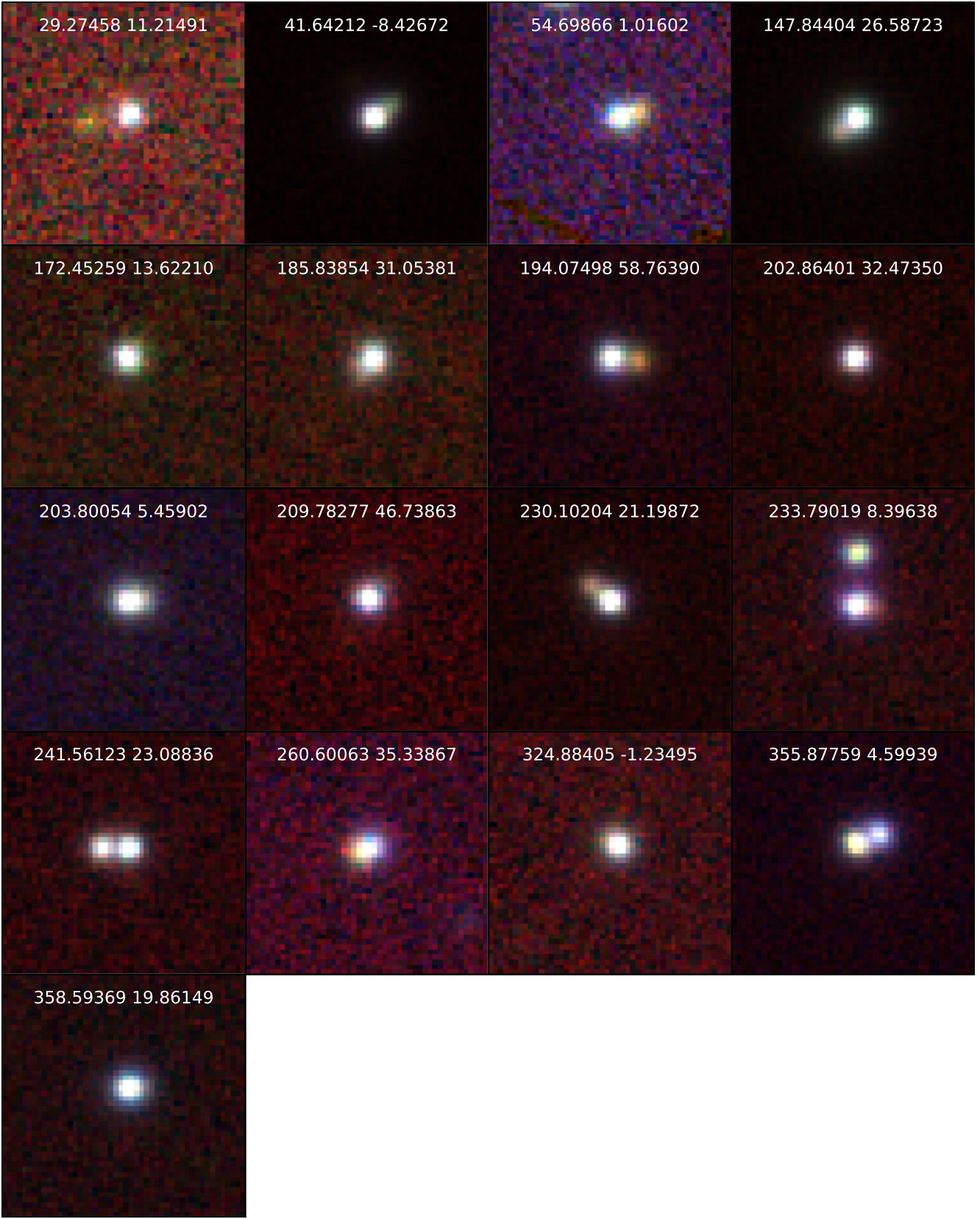}
\caption{Pan-STARRS $g,i,y$ cutouts of 17 objects listed in Table \ref{cand.tab}, which are either resolved in Gaia EDR3 into separate components, or are suspected to be double based on their images. North is up, east to the left. Each image is $12\arcsec$ on a side.} \label{cutouts.fig}
\end{figure*}

There are 44 sources (out of 47) with definitely resolved and suggested companions in Table \ref{cand.tab}, which is almost half the expected number of genuinely perturbed cases. Three sources are included on account of their outstanding properties as radio-loud ICRF quasars and blazars. Four objects stands out as known doubly or multiply imaged gravitational lens. A relative motion of the lens and the quasar and a caustic crossing may cause shifts and sudden jumps in the measured position of the source \citep{2004A&A...416...19T}. Could the other suspected and resolved cases of multiplicity be new strong lenses? High-resolution imaging is needed to answer this question. The appearance of companions suggests that they are either double AGNs or lensed images. Although the probability for each source to have a chance neighbor within $3\arcsec$ is only 0.0076, and within $2\arcsec$ is 0.0042, see Eq. \ref{phi.eq}, the estimated rate of companions in the high completeness sample, $44/152=0.29$, indicates that the occurrence of perturbed proper motions is strongly correlated with close multiplicity. A panorama composition of Pan-STARRS images of the high completeness selected sample quasars is given in Figure \ref{cutouts.fig}.
 
As mentioned earlier, the majority of the reviewed sample do not display any obvious signs of morphological irregularity, which is partly explained by the expected rate of statistical outliers and partly explainable by angular resolution limitations. We note, however, the presence of two radio-loud ICRF3 sources in Table \ref{cand.tab} without any optical companions, but with previously reported radio-optical position offsets and, in one case at least, with significant offsets between the radio bands determinations between the three components of ICRF3. For these sources, jet variability and extent may be responsible for the appearance of proper motion.
 
\section{Near-neighbor distance analysis} \label{nn.sec}
The nearest-neighbor distance statistic is a powerful method of analysis and detection of populations with non-random spatial distributions \citep{1954Ecol...35..445C}. It belongs to the class of first-order statistics of a sample. In this application, we consider the angular separations between each of an initial cleaned sample of $549,517$ MIRAGN/Gaia quasars with measured proper motions and all their neighbors within an $11\arcsec$ radius. This outer limit is small enough to generate a manageable sample of neighbors and large enough to collect a large sample of optical companions at separations beyond several arcseconds where physically double AGNs are unlikely to appear. The total number of Gaia EDR3 sources thus selected (including the $549,517$ primary targets and all other Gaia entries within $11\arcsec$) is $748,642$. We find that $113,240$ quasars have a total of $199,125$ companions within $11\arcsec$, which means that the majority of quasars with a companion have more than one companion although most of the quasars ($436,277$) do not have companions at all. This is the first sign of a strong clustering property. The number of targets without companions can be used to estimate an important scaling parameter. For a purely random positioning of sources on the celestial globe, the probability of not having at least one companion within an angular radius $r$ in radians is 
\eb
P_{\rm empty}=\left( 1-\frac{\pi r^2}{4\pi}\right)^N
\label{pempty.eq}
\ee
\noindent where $N$ is the total number of sources, which defines the effective number density. This probability is accurately estimated as the ratio of targets without any companions to the total number of targets. Using the numbers above and $r=11\arcsec$, we obtain $N=3.246\times 10^8$. The effective number of sources is much smaller than the number of sources in Gaia EDR3 (1.8 billion) because the footprint of MIRAGN does not include the Galactic equator belt where most of Gaia stars are located. 
 
Using this estimate and Equation~\ref{pempty.eq}, the expected rate of empty (companion-free) circles can be easily computed for any $r<11\arcsec$. For each of the $113,240$ targets, the distance to the nearest neighbor $r_{\rm nn}$ is computed. The sample analog of $P_{\rm empty}(r)$ is the ratio of the number of targets with $r_{\rm nn}>r$ to the total number of targets, which is denoted
$\Phi_{\rm obs}$. Alternatively, the rate of targets with at least one companion within $r$, $N[r_{\rm nn}<r]/N_{\rm targets}$ can be compared with the expected rate,which is
\eb
\Phi_{\rm exp}=P[r_{\rm nn}<r]=1-{\rm dex}(N\,\log(1-r^2/4)).
\label{phi.eq}
\ee
 
\begin{figure}[htbp]
  \centering
\epsscale{1}
  \plottwo{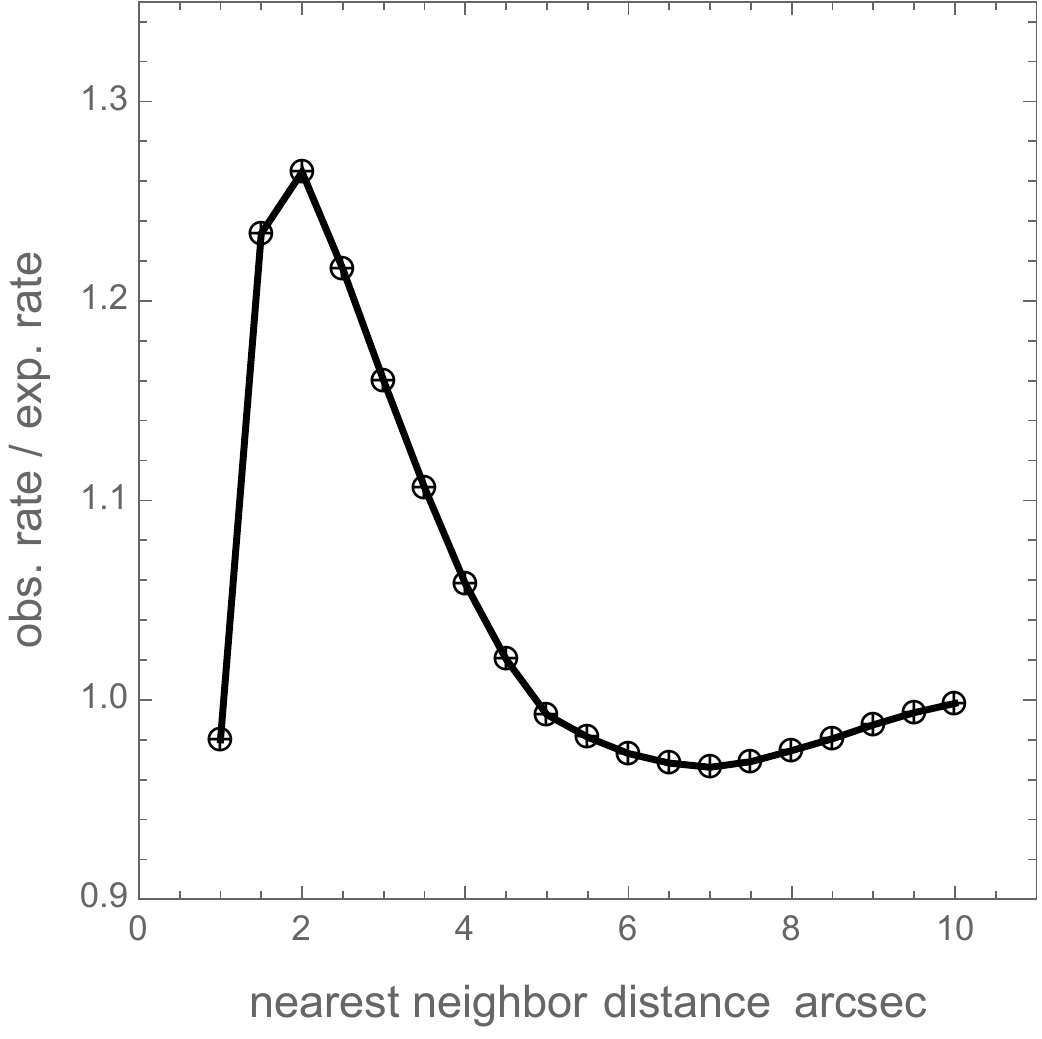}{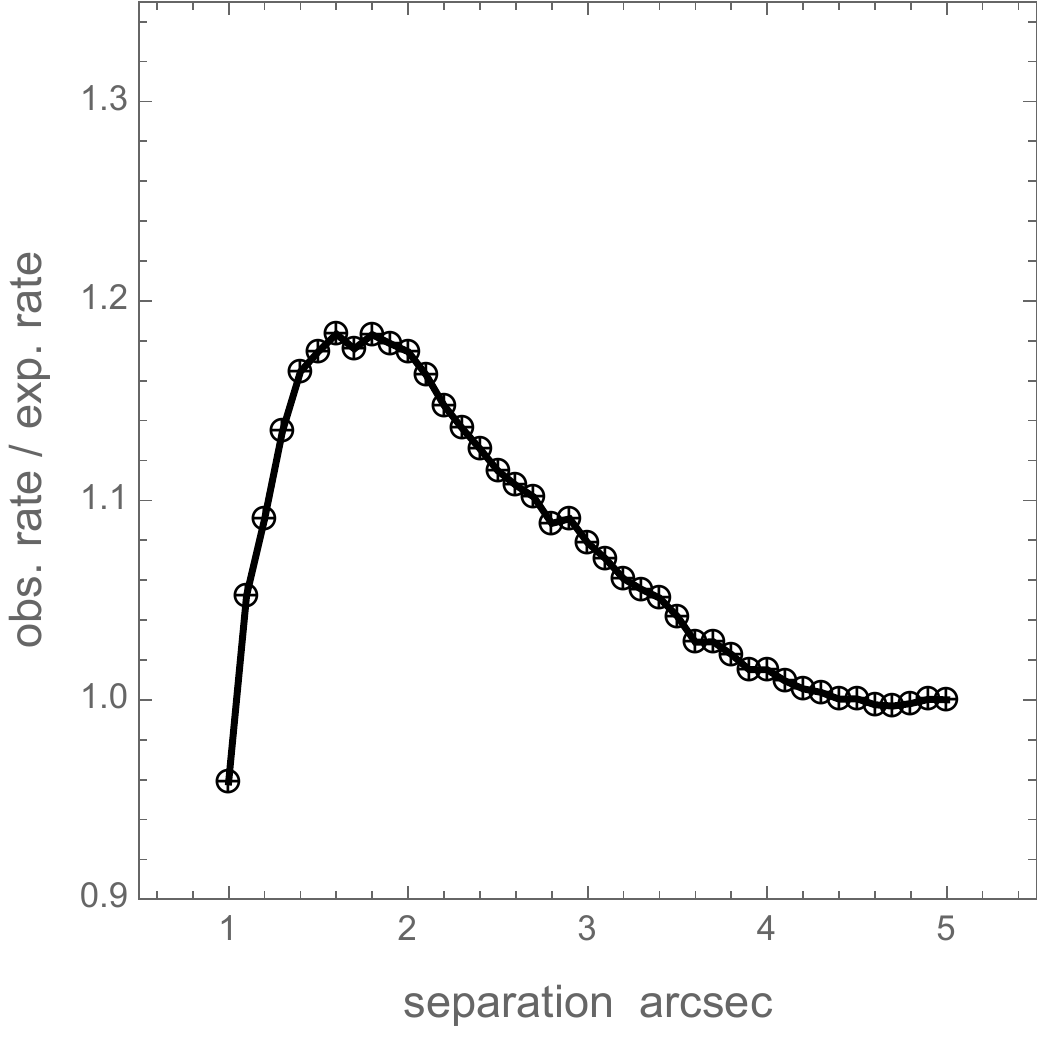}
\caption{Left: the ratio of observed rate and expected rate  of nearest-neighbor distances in a sample of $113,240$ quasars with at least one companion source within $11\arcsec$ in Gaia DR3.
Right: the ratio of observed rate and expected rate of angular distances in a sample of $11,446$ quasars with only one companion source within $5\arcsec$ in Gaia DR3.
\label{nnrate.fig}}
\end{figure}

Figure \ref{nnrate.fig}, left panel, shows the ratio of the observed and expected rates of nearest-neighbor distances $\Phi_{\rm obs}/\Phi_{\rm exp}$ for a grid of $r$. The observed rate of close companions is higher than the expectation for separations within $4\arcsec$ except the smallest value $r=1\arcsec$. At this small separation, the deficit of neighbors is caused by a drop in catalog completeness for pair separations below $\sim1\farcs5$, as shown in \citet{2021A&A...649A...5F}. It should be noted that the pairwise distance statistics used by \citet{2021A&A...649A...5F} is essentially different from the nearest neighbor distance statistics, being possibly perturbed at small separations, especially if the bin size is comparable to the separation. At separations greater than $1\farcs5$, we find a significant excess in this large sample of companions. The excess rate equals 23\% of the expected value at $r=1\farcs5$, peaks at 26\% for $r=2\farcs0$, and slowly declines toward $4\arcsec$. The expected rate can be used to estimate the number of quasars with stellar optical companions, assuming that the nearest companions at $11\arcsec$ are all stars. Given the total number of primary quasars $N_{\rm prim}=549\,517$ and the estimated probability $\Phi_{\rm est}$ in Eq. \ref{phi.eq}, the total number of quasars in the general sample with genuine companions should then be 4176 for a limiting $r_{\rm nn}=1\farcs5$ and 6511 for $2\farcs0$. The corresponding observed rates are 0.4\% and 0.8\%. This is lower that the estimated lower-bound rate of physical dual (binary) AGNs of about 1\% or greater in the literature \citep[e.g.,][]{2020ApJ...904...23K}. Our estimates are expected to be underestimated due to the limited angular resolution capabilities of Gaia at small separations and faint magnitudes, where most of double AGNs are hidden.

The nearest-neighbor statistic suffers from the natural variations of the local density of objects. This is true for our initial sample despite the avoidance of the Galactic plane and additional cuts designed to eliminate confused sources in crowded areas. To illustrate this point, we note that, while 79\% of the initial sample do not have neighbors within $11\arcsec$, $0.0011\%$ of the sample have 16 companions within the same radius. These occasional overcrowded regions contribute to the peak we see in Fig. \ref{nnrate.fig}, left, therefore, we cannot conclude that the excess at small separations is entirely due to physical double AGNs. To obtain a more accurate estimate of the latter, we consider only those sources that have only one neighbor within $r_{\rm lim}$, which eliminates high-density areas. The previous analysis of nearest-neighbor distance probability does not apply anymore because the statistic is strongly biased toward wider separations by the elimination of sources with multiple companions. Given a sufficiently small $r_{\rm lim}$, the relative rate of companions within concentric circles of smaller radii should still reflect an excess of tight physical companions, if any. Choosing $r_{\rm lim}=5\arcsec$, we performed this source count for the corresponding sample of $11,446$ quasars. The results are shown at a higher resolution in separation in \ref{nnrate.fig}, right panel. Assuming a uniform probability of companion location, the expected rate of single companions within a circle of radius $r$ is proportional to $r^2/r_{\rm lim}^2$. The ratio of the observed counts to the expected rate in this figure shows a strong peak at $\sim 1.8\arcsec$ and a steep drop toward $1\arcsec$ of undoubtedly instrumental origin. If all of the companions at $5\arcsec$ are optical, the excess rate of physical companions relative to the rate of optical pairs is roughly 0.18 in the separation range $[1.5,2.0]\arcsec$. Extending this estimation based on Eq. \ref{phi.eq} and Fig. \ref{nnrate.fig} to the entire sample  of MIRAGN quasars at all separations yields a rate of genuine double (or dual) AGNs and quasars of at least 0.20\%. Most of them are likely to be hidden at separations smaller than $1\arcsec$. We also note that this analysis is only sensitive to sufficiently bright companions.

\section{Chance alignment with stars and weak lensing}
Using our estimates in Sect \ref{nn.sec}, we estimate that the sources in the general MIRAGN/EDR3 sample are expected to have 14 optical companions within $0.1\arcsec$ and 1445 within $1.0\arcsec$. The rate of foreground neighbors is as high as 0.003 per square arcsecond on average for the entire sky---however, it drops down to approximately 0.001 for the low-density areas away from the Galactic plane covered by SDSS. A foreground object causes gravitational deflection of light rays from a distant source, and the proper motion of the lens generates a time-variable component of the astrometric displacement. The Einstein ring radius for a star of one solar mass at 1 kpc distance is only 2.85 mas, so strong lensing by a chance star is unlikely to be present, even taking into account the numerous stars not visible to Gaia. The weak lensing effect for realistic quasar-star configurations is typically quite small. Using formulae from \citep{1995A&A...294..287H,2000ApJ...534..213D,2016MNRAS.460.2025K}, we calculate that the proper motion of a quasar induced by a solar mass lens at a distance of 10 pc with an impact angle of $1\arcsec$ and a proper motion of 10 \masyr\ maximizes at the closest approach with an amplitude of 8.12 $\mu$as yr$^{-1}$\footnote{For such nearby lenses, the parallax may generate a larger apparent motion of the source than the proper motion}. The effect is roughly inversely proportional to the impact separation squared and proportional to the lens' proper motion, so that faster moving closer neighbors can, in principle, generate measurable quasar proper motions. Such events should be extremely rare, however \citep{1997AJ....114.1508H}. 

\section{Discussion and Conclusions}
Using a large source sample of 551,482 high-fidelity quasars with proper motion estimates, we estimated a rate of 0.17\% of genuinely perturbed proper motion in excess of the expected rate of statistical outliers (Section~\ref{rate.sec}). A set of photometric and astrometric filters has been applied to the initial, much larger working sample of MIRAGN/EDR3 sources to minimize the impact of the small population of stellar interlopers, so that this estimate is considered to be reliable. Furthermore, we derived a general correction factor of 1.06 that we applied to all formal proper motion errors, as explained in Sect. 
\ref{section: Methodology}, which brings the distribution of normalized total proper motions in line with the Rayleigh distribution expected for random statistical errors (Figure \ref{fig:pmchi}). In fact, we do not know if the difference between these distributions is caused by generally underestimated uncertainties in Gaia EDR3. A less likely proposition is that AGNs have a common source of astrometric perturbation related to their time-dependent structure of the order of 25 $\mu$as yr$^{-1}$. This value, however, is consistent with the bounding value of the fitted covariance function at zero separation \citep{2021A&A...649A...2L}. Our chosen conservative approach to this problem implies that the estimated rate of perturbed proper motions is underestimated.

Starting with a null hypothesis that the proper motion excess is primarily of instrumental rather then astrophysical origin, we want to determine which circumstances may contribute to the emergence of perturbed reference quasars. Noting that the excess flux ratio parameter  \texttt{phot\_bp\_rp\_excess\_factor} provided in EDR3 is strongly correlated with redshift, we find that nearby reference quasars are associated with resolved host galaxies in Pan-STARRS images, which are often asymmetric and complex. The Gaia astrometric pipeline, designed for unresolved sources (i.e., stars), cannot handle these extended components accurately, which leads to additional unexplained variance (i.e., larger \texttt{astrometric\_excess\_noise}). 

We produced a sample of 47 AGNs with bona fide apparent proper motions, and constructed a control sample of 470 AGNs, matched in Gaia $G$ magnitude and S/N, with no evidence for proper motion, finding that several highly significant ($p < 0.001$) differences that collectively implicate double and multi-sources as being responsible for apparent proper motions.

To explore this, a small test sample of 152 bona fide quasars, selected to have a high completeness of sources with apparently significant proper motions and spectroscopically determined redshifts greater than 0.5 in the footprint of the Pan-STARRS $3\pi$ survey was investigated on a source-by-source basis. While this visual inspection and extensive literature search do not reveal any peculiarities for two-thirds of the sample, at least 44 objects (29\%) have companions within $2\farcs5$. In the concordance $\Lambda$CDM cosmology, $1\arcsec$ on the sky spans $\sim6$~kpc in the rest frame of a source at $z=0.5$, and $\sim8$~kpc at $z=1.0$, so these companions can be true dual AGNs. Radio-loud quasars from the ICRF3 catalog with previously reported radio-optical position offsets also appear in the high completeness sample. The boosted proper motion may be correlated with the statistically significant radio-optical and multiwavelength radio offsets \citep{2021A&A...652A..87L} if they have a common origin in addition to the alignment with relativistic radio jets \citep{2021AA...651A..64L}. 

Numerically, most of the sources with perturbed proper motions have low  error-normalized total proper motions $\chi=\mu/\sigma_\mu$ values, between 2.2 and 5 (Figure \ref{rate.fig}), i.e., they are marginally significant in Gaia EDR3. The expected number of genuinely perturbed outliers in the high completeness sample of 152 sources is 99. With 47 sources listed in Table \ref{cand.tab}, we have identified plausible peculiarities for almost half of them. However, we have likely missed quite a few sources with companions closer than about $1\arcsec$ due to the limited angular resolution of Gaia, and the correlation between proper motion perturbation and multiplicity is likely to be stronger. A nearest-neighbor distance statistical analysis (Sect. \ref{nn.sec}) of a much larger, all-sky sample of bona fide MIRAGN/Gaia quasars reveals that there is a clear excess of close companions to AGNs peaking at $26\%$ relative to the expected rate of chance stellar neighbors for $r=2\arcsec$. In absolute terms, the approximately estimated rate of surplus companions at small separations is 0.2\% or higher for the entire sample of MIRAGN quasars. The extra close companions are likely to be genuine double ( dual) AGNs, strongly lensed quasars, and galactic mergers at small redshifts. Most of such objects remain unresolved and hidden, with separations less than $1\arcsec$. Follow-up imaging observations with the HST, high-resolution ground-based facilities, and spectroscopy \citep[e.g., looking for split \text{[\ion{O}{3}]} emission lines; see, however, the limitations of this method from gas outflow kinematics,][]{2012ApJ...745...67F} should confirm the presence of yet unresolved double AGNs. 

In view of the strong correlation of excess proper motions and physical and optical multiplicity for quasars and AGNs, what is the mechanism of astrometric perturbation? Four objects in the high completeness sample are known multiply imaged strong gravitational lens. The relative motion of the lens galaxy and the source quasar is greatly magnified in such cases, which can conceivably lead to an apparent motion of the split images. These cases are too rare compared to the estimated rate of proper motion quasars, however.
A more likely scenario for quasar apparent proper motions is the VIM effect. The origin of this effect goes back to the methodology of astrometric estimation at the photon counts level. VIM is a powerful emerging method for detection of sub-kiloparsec double and dislodged AGNs and lenses \citep{2020ApJ...888...73H}. In the simplest models (such as the one implemented for the Kepler mission data processing), the center of the image is computed as the first moment of unweighted pixel fluxes within a fixed mask (also called aperture or window). This first-moment centroid is very sensitive to the presence of a neighbor within the aperture. Even if only a small part of the neighbor's image is present from the wing of the point spread function (PSF), it shifts the estimated photocenter toward the neighbor. More advanced and sophisticated methods of weighted moment estimation or PSF fitting are employed in large astrometric projects. These are more robust to perturbations from neighbors, which are mostly confined to close and unresolved sources. Still, an astrometric displacement is inevitable with a magnitude depending on the PSF width and shape, the separation, and the relative flux in the instrument passband. The shift hardly matters for regular, constant sources (unless we are concerned with absolute positions of CRF sources), but variability of the source or one of its companions makes the photocenter vary along the line connecting the sources, possibly producing a bogus proper motion. The astrometric trajectory may reflect the light curve with extreme fidelity in such cases \citep{2016ApJS..224...19M}. For the Gaia mission, astrometric estimation is implemented in the broad $G$ band using an exhaustively calibrated one-dimensional line spread function \citep[LSF,][]{2021A&A...649A..11R}. It is composed of a fixed mean profile for point-like sources plus a number of calibrated basis terms represented by spline-interpolation functions, which include both symmetric and asymmetric components. The full width at half maximum of the main component is not less than $1.8\arcsec$ \citep{2021A&A...649A..11R}, which defines the effective angular resolution of Gaia. The calibrated basis terms are functions of time, also subject to jumps at the decontamination events. 

Quasars are intrinsically variable optical sources, so they must be burdened by VIM perturbations when found in tight physical or optical pairs. A dedicated study of ICRF3 radio-loud quasars shows that the typical rms amplitude in the optical bands is about 0.1 mag \citep{2021AJ....162...21B} but some extreme objects (blazars) may have variations of 2 mag. The amplitude of variability tends to appreciably decline with redshift, which is explained as a time dilation effect in combination with the red power spectrum. The latter implies that VIM should mostly affect the proper motions of nearby AGNs. A decisive test for this hypothesis would be an anti-correlation between redshift and proper motion excess $\chi$. Unfortunately, AGNs with $z<0.5$ tend to be associated with extended images of host galaxies. The LSF fitting is not tuned to extended sources, resulting in systematically underestimated formal errors of Gaia astrometry despite the post-fit scaling of formal errors by the actual rms residuals. This verification of the VIM scenario hinges on the yet unexplored issue of how well the Renormalised Unit Weight Error \citep[RUWE;][]{2021A&A...649A...2L} parameter in Gaia~EDR3 captures the astrometric degradation caused by the galaxies surrounding nearby AGNs.


\begin{acknowledgements}
The authors thank the anonymous referee for their helpful review that improved this work. 

This research made use of Astropy,\footnote{http://www.astropy.org} a community-developed core Python package for Astronomy \citep{2013A&A...558A..33A, 2018AJ....156..123A}, and \textsc{topcat} \citep{2005ASPC..347...29T}

Funding for the Sloan Digital Sky Survey IV has been provided by the Alfred P. Sloan Foundation, the U.S. Department of Energy Office of Science, and the Participating Institutions. 

SDSS-IV acknowledges support and resources from the Center for High Performance Computing  at the University of Utah. The SDSS website is www.sdss.org.

SDSS-IV is managed by the Astrophysical Research Consortium for the Participating Institutions of the SDSS Collaboration including the Brazilian Participation Group, the Carnegie Institution for Science, Carnegie Mellon University, Center for Astrophysics | Harvard \& Smithsonian, the Chilean Participation Group, the French Participation Group, Instituto de Astrof\'isica de Canarias, The Johns Hopkins University, Kavli Institute for the Physics and Mathematics of the Universe (IPMU) / University of Tokyo, the Korean Participation Group, Lawrence Berkeley National Laboratory, Leibniz Institut f\"ur Astrophysik Potsdam (AIP),  Max-Planck-Institut f\"ur Astronomie (MPIA Heidelberg), Max-Planck-Institut f\"ur Astrophysik (MPA Garching), Max-Planck-Institut f\"ur Extraterrestrische Physik (MPE), National Astronomical Observatories of China,New Mexico State University, New York University, University of Notre Dame, Observat\'ario Nacional / MCTI, The Ohio State University, Pennsylvania State University, Shanghai Astronomical Observatory, United Kingdom Participation Group, Universidad Nacional Aut\'onoma de M\'exico, University of Arizona, University of Colorado Boulder, University of Oxford, University of Portsmouth, University of Utah, University of Virginia, University of Washington, University of Wisconsin, Vanderbilt University, and Yale University.

The Pan-STARRS1 Surveys (PS1) and the PS1 public science archive have been made possible through contributions by the Institute for Astronomy, the University of Hawaii, the Pan-STARRS Project Office, the Max-Planck Society and its participating institutes, the Max Planck Institute for Astronomy, Heidelberg and the Max Planck Institute for Extraterrestrial Physics, Garching, The Johns Hopkins University, Durham University, the University of Edinburgh, the Queen's University Belfast, the Harvard-Smithsonian Center for Astrophysics, the Las Cumbres Observatory Global Telescope Network Incorporated, the National Central University of Taiwan, the Space Telescope Science Institute, the National Aeronautics and Space Administration under Grant No. NNX08AR22G issued through the Planetary Science Division of the NASA Science Mission Directorate, the National Science Foundation Grant No. AST-1238877, the University of Maryland, Eotvos Lorand University (ELTE), the Los Alamos National Laboratory, and the Gordon and Betty Moore Foundation.

\end{acknowledgements}
 
\facilities{Gaia, WISE, Sloan, PS1}

\software{Astropy \citep{2013A&A...558A..33A, 2018AJ....156..123A}, \textsc{topcat} \citep{2005ASPC..347...29T}}
 
\bibliography{manuscript}
\bibliographystyle{aasjournal}

\appendix
Table 3 provides the list of objects in the high-reliability sample of quasars with excess proper motions described in Section 3.1.
\begin{deluxetable}{crccr}
\tablehead{\colhead{MIRAGN ObjID} & \colhead{Gaia source ID} & \colhead{R.A.} & \colhead{decl.} & \colhead{$\chi$}\\ [-0.2cm]
\colhead{ } & \colhead{ } & \colhead{($\arcdeg$)} & \colhead{($\arcdeg$)} & \colhead{$\mathrm{}$}}
\startdata
J001813.29$+$361058.7 & 2876570633012282112 & 004.55543 & $+$36.18294 & 7.19 \\
J021425.92$-$024254.4 & 2493291050051966592 & 033.60807 & $-$02.71511 & 5.76 \\
J074922.97$+$225511.9 & 675266509307714304 & 117.34569 & $+$22.91994 & 6.16 \\
J075824.27$+$145752.5 & 654600574086014976 & 119.60110 & $+$14.96457 & 10.14 \\
J080559.23$+$490742.0 & 934597390554520704 & 121.49682 & $+$49.12836 & 13.49 \\
J081331.28$+$254503.1 & 682614034415644032 & 123.38030 & $+$25.75086 & 8.37 \\
J083531.06$+$252808.7 & 702664277488240512 & 128.87949 & $+$25.46910 & 5.49 \\
J090933.50$+$425346.5 & 817131306319108096 & 137.38957 & $+$42.89624 & 5.52 \\
J091831.59$+$110653.1 & 592633679290265984 & 139.63161 & $+$11.11474 & 5.14 \\
J095738.17$+$552257.7 & 1045124392483104512 & 149.40910 & $+$55.38271 & 10.00 \\
J110720.43$+$152230.0 & 3969511549236225792 & 166.83511 & $+$15.37498 & 5.06 \\
J111411.89$+$522935.9 & 842349640591293824 & 168.54955 & $+$52.49330 & 8.79 \\
J122016.86$+$112628.2 & 3907389726382580352 & 185.07031 & $+$11.44114 & 5.71 \\
J122321.24$+$310313.8 & 4015077132157656832 & 185.83854 & $+$31.05382 & 5.06 \\
J122733.68$+$435519.8 & 1535614154516228736 & 186.89032 & $+$43.92216 & 8.30 \\
J123143.56$+$284749.7 & 4010804292533346176 & 187.93155 & $+$28.79716 & 5.78 \\
J125617.97$+$584550.0 & 1578826237094189184 & 194.07498 & $+$58.76390 & 5.43 \\
J125631.35$+$330253.1 & 1515468558874746368 & 194.13063 & $+$33.04802 & 5.09 \\
J132559.51$+$144110.4 & 3743298915995865856 & 201.49802 & $+$14.68623 & 28.01 \\
J133127.36$+$322824.6 & 1469088306556900864 & 202.86401 & $+$32.47350 & 6.91 \\
J135846.74$+$075150.4 & 3721019099565323392 & 209.69471 & $+$07.86402 & 5.11 \\
J140901.89$+$294633.5 & 1453110272301974016 & 212.25787 & $+$29.77597 & 9.83 \\
J142111.91$+$353829.3 & 1480789068781698944 & 215.29963 & $+$35.64147 & 21.68 \\
J143100.00$-$013141.7 & 3649591907942645888 & 217.75000 & $-$01.52827 & 9.41 \\
J143454.72$+$285624.4 & 1281111744223168896 & 218.72801 & $+$28.94014 & 10.85 \\
J144034.78$+$441520.5 & 1493598207446623104 & 220.14495 & $+$44.25570 & 8.40 \\
J144444.64$+$223902.5 & 1241776376438184704 & 221.18606 & $+$22.65075 & 5.32 \\
J145439.57$+$334904.2 & 1289696627934152576 & 223.66490 & $+$33.81783 & 6.85 \\
J152024.51$+$211155.4 & 1214124621072141952 & 230.10204 & $+$21.19872 & 8.83 \\
J152902.83$+$384103.1 & 1387903838295453440 & 232.26182 & $+$38.68421 & 5.60 \\
J153038.05$+$545631.7 & 1601059835380727552 & 232.65863 & $+$54.94208 & 5.05 \\
J154049.49$+$144746.0 & 1194474282403600128 & 235.20621 & $+$14.79608 & 9.96 \\
J160614.71$+$230518.0 & 1206719375899944320 & 241.56123 & $+$23.08836 & 6.06 \\
J160829.24$+$271626.9 & 1315986917323899392 & 242.12180 & $+$27.27410 & 12.40 \\
J161827.73$+$505817.6 & 1424429069110587904 & 244.61552 & $+$50.97156 & 17.58 \\
J162454.24$+$495312.3 & 1423345466042177920 & 246.22603 & $+$49.88674 & 7.46 \\
J162525.49$+$340716.9 & 1326105546739413504 & 246.35626 & $+$34.12136 & 8.17 \\
J165043.44$+$425148.9 & 1356637374030072832 & 252.68102 & $+$42.86372 & 5.92 \\
J171836.60$+$381835.3 & 1340677786658074240 & 259.65253 & $+$38.30986 & 5.92 \\
J172224.15$+$352019.2 & 1336607605067137664 & 260.60063 & $+$35.33867 & 5.69 \\
J172758.48$+$564419.4 & 1422294500428597120 & 261.99371 & $+$56.73868 & 8.72 \\
J173330.84$+$552030.8 & 1420997626464142464 & 263.37852 & $+$55.34190 & 15.12 \\
J213932.19$-$011405.4 & 2674574094834071936 & 324.88405 & $-$01.23495 & 9.02 \\
J222634.21$-$011850.9 & 2629900009684328192 & 336.64259 & $-$01.31415 & 7.36 \\
J234330.59$+$043557.9 & 2743952423847631616 & 355.87760 & $+$04.59939 & 7.15 \\
J235422.48$+$195141.3 & 2822517236937352832 & 358.59369 & $+$19.86149 & 7.52 \\
J235701.73$-$022545.7 & 2449007639424063616 & 359.25722 & $-$02.42934 & 9.09
\enddata
\caption{High-reliability proper motion sample. \label{tab: high reliability sample}}
\end{deluxetable}

\end{document}